\documentclass{emulateapj}
\usepackage{amsmath}
\usepackage{amssymb}
\usepackage{ulem}
\usepackage{longtable}
\usepackage{graphicx}
\usepackage{epstopdf}
\usepackage{booktabs}
\usepackage{natbib}
\usepackage[colorlinks, citecolor=blue]{hyperref}

\shorttitle{} \shortauthors{}

\begin{document}

\title{The Origin of the Prompt Emission for Short GRB 170817A: Photosphere
Emission or Synchrotron Emission?}
\author{Yan-Zhi Meng\altaffilmark{1,2}, Jin-Jun Geng\altaffilmark{3,4},
Bin-Bin Zhang\altaffilmark{3,4}, Jun-Jie Wei\altaffilmark{1}, Di Xiao%
\altaffilmark{3,4}, Liang-Duan Liu\altaffilmark{3,4}, He Gao\altaffilmark{5}%
, Xue-Feng Wu\altaffilmark{1}, En-Wei Liang\altaffilmark{6}, Yong-Feng Huang%
\altaffilmark{3,4}, Zi-Gao Dai\altaffilmark{3,4}, Bing Zhang\altaffilmark{7}}

\begin{abstract}
The first gravitational-wave event from the merger of a binary neutron star
system (GW170817) was detected recently. The associated short gamma-ray
burst (GRB 170817A) has a low isotropic luminosity ($\sim 10^{47}$ erg s$%
^{-1}$) and a peak energy $E_{p}\sim 145$ keV during the initial main
emission between -0.3 and 0.4 s. The origin of this short GRB is still under
debate, but a plausible interpretation is that it is due to the off-axis
emission from a structured jet. We consider two possibilities. First, since
the best-fit spectral model for the main pulse of GRB 170817A is a cutoff
power law with a hard low-energy photon index ($\alpha
=-0.62_{-0.54}^{+0.49} $), we consider an off-axis photosphere model. We
develop a theory of photosphere emission in a structured jet and find that
such a model can reproduce a low-energy photon index that is softer than a
blackbody through enhancing high-latitude emission. The model can naturally
account for the observed spectrum. The best-fit Lorentz factor along the
line of sight is $\sim 20$, which demands that there is a significant delay
between the merger and jet launching. Alternatively, we consider that the
emission is produced via synchrotron radiation in an optically thin region
in an expanding jet with decreasing magnetic fields. This model does not
require a delay of jet launching but demands a larger bulk Lorentz factor
along the line of sight. We perform Markov Chain Monte Carlo fitting to the
data within the framework of both models and obtain good fitting results in
both cases.
\end{abstract}


\affil{\altaffilmark{1}Purple Mountain Observatory, Chinese Academy
of Sciences, Nanjing 210008, China; xfwu@pmo.ac.cn}
\affil{\altaffilmark{2}University of Chinese Academy of Sciences,
Beijing 100049, China}
\affil{\altaffilmark{3}School of Astronomy
and Space Science, Nanjing University, Nanjing 210093, China;
gengjinjun@nju.edu.cn, zhang.grb@gmail.com }
\affil{\altaffilmark{4}Key Laboratory of Modern Astronomy and
Astrophysics (Nanjing University), Ministry of Education, China}
\affil{\altaffilmark{5}Department of Astronomy, Beijing Normal
University, Beijing 100875, China}
\affil{\altaffilmark{6}Department
of Physics and GXU-NAOC Center for Astrophysics and Space Sciences,
Guangxi University, Nanning 530004, China}
\affil{\altaffilmark{7}Department of Physics and Astronomy,
University of Nevada, Las Vegas, NV 89154, USA }

\section{INTRODUCTION}

\label{sec:intro}

Recently, the first joint detection of gravitational wave (GW) event
(GW170817;~\citealt{LIGO2017a}) and short gamma-ray burst (GRB 170817A;~%
\citealt{LIGO2017b,Connau2017,Gold2017a,Sav2017}) confirmed the hypothesis
that mergers of the double neutron stars (NS$-$NS) are the progenitor
systems of short gamma-ray bursts (SGRBs; %
\citealt{Eich1989,Nara1992,Moch1993,Nakar2007,Berger2014}). Follow-up
electromagnetic observations revealed a host galaxy of GRB 170817A at a
distance of $\sim 40$~Mpc \citep{Coul2017}, as well as broad-band emission %
\citep{LIGO2017c}. The isotropic-equivalent energy of GRB 170817A is $\sim
5\times 10^{46}$~erg \citep{Gold2017,ZhangBB18b}, which is much smaller than
that of a typical SGRB ($10^{50}$~erg).

Previous observations of short GRB jet breaks suggested that the half
opening angle of a SGRB jet is $\leq 20^{\circ }$ \citep[e.g.][]{Fong2015}.
On the other hand, the GW signals are essentially isotropic, so the
detection rate of a GW event associated with an on-axis burst should be
quite low for binary NS mergers. However, the simultaneous detection of GRB
170817A and GW170817 indicates that the rate for such similar events is
actually high \citep{ZhangBB18b}. Such a high rate implies that the jet may
be structured, with an angle-dependent luminosity and bulk Lorentz factor
outside an uniform core, rather than a simple `top-hat' form with a sharp
edge \citep{Gran2017}. Emission from such a structured jet could thus be
seen by an off-axis observer with a large viewing angle (e.g., %
\citealt{Jin2017,Lamb2017,Lazza2017,Xiao2017,Kath2018}). The low
isotropic luminosity ($\sim $ $10^{47}$ erg s$^{-1}$) of the prompt
emission for GRB 170817A \citep{Gold2017,ZhangBB18b} does support
this suggestion. A
structured jet has also been favored by other recent theoretical %
\citep[e.g.,][]{Sap2014} and numerical %
\citep[e.g.,][]{Aloy2005,Tche2008,Komi2010,Murgu2017} studies within the NS$%
- $NS merger context. As the jet breaks out of the neutron-rich
\textquotedblleft dynamical ejecta\textquotedblright\ ejected during the
merger \citep[e.g.,][]{Hoto2013,Ross2013}, some \textquotedblleft lateral
structure\textquotedblright\ would be developed, which has a lower
luminosity than the on-axis relativistic jet.

The prompt emission for GRB 170817A is shown to have two temporal
components: a main pulse and a weak tail. The main pulse ($-0.26$ to $0.57$
s) spectrum is well fitted by the cutoff power-law model with the low-energy
photon index $\alpha =-0.61_{-0.60}^{+0.34}$, while the weak tail ($0.95$---$%
1.79$ s), with $\sim $ 1/3 of the fluence of the main pulse, is well fitted
by a blackbody model \citep[][see also \cite{Gold2017a}]{ZhangBB18b}.

The physical origin of the prompt emission of GRB 170817A is unknown. The
exponential cutoff on the high-energy end and the relatively hard low-energy
photon index (i.e., $\alpha =-0.61$ for the time interval between $-0.26$
and $0.57$ s) for the main pulse and the dominated blackbody in the weak
tail may support a possible photospheric origin of the emission %
\citep[e.g.,][]{Good1986,Pac1986,Abra1991,Thom1994,Me2000,Me2002,Ry2004,Ry2005,Ree2005,Abdo2009,Pe2011,Lund2013,Deng2014,Begue2015,Gao2015,Pe2015}%
. On the other hand, the $\alpha $ index is also consistent with the
typical $\alpha =-2/3$ segment of synchrotron radiation
\citep{rybicki79}. It is therefore interesting to perform detailed
modeling of the prompt emission using both photospheric and
synchrotron models, especially within the framework of an off-axis
structured jet.

This paper is organized as follows. In Section \ref{sec:mod}, we develop a
model of off-axis photosphere emission from a structured jet. Then we apply
this model to perform a Markov Chain Monte Carlo (MCMC) fitting to the
spectrum of the main pulse of GRB 170817A in Section \ref{sec:fit}. In
Section \ref{sec:syn}, we apply the MCMC technique to fit the same spectrum
using the synchrotron model. Section \ref{sec:dis} presents some discussions
and the conclusions are drawn in Section \ref{sec:con}.

\section{Off-axis photosphere model in a structured jet}

\label{sec:mod}

In this section, we present the calculation of the time-integrated
photospheric emission spectrum from a structured jet observed by an off-axis
observer.

\subsection{Jet Structure}

\begin{figure}[bh]
\label{Fig_1} \centering\includegraphics[angle=0,scale=0.27]{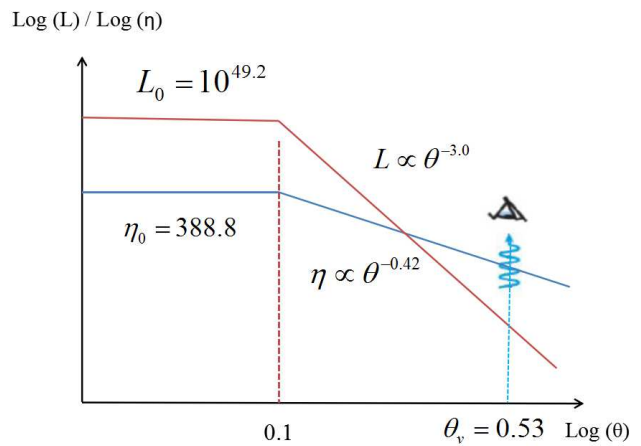}
\caption{Jet structure and viewing angle for our photosphere model fitting
of the main pulse spectrum ($-0.3$ s to $0.4$ s) of GRB 170817A. For our
photosphere model fitting in Section 3, the best-fit values are $L_{0}\sim
10^{49.16}$ erg s$^{-1}$, $\protect\theta _{c,L}\sim 0.11$ rad and $q\sim $ $%
2.99$ for the angular profile of luminosity, $\protect\eta _{0}\sim 388.82$,
$\protect\theta _{c,\Gamma }\sim 0.11$ rad and $p$ $\sim
0.42_{-0.07}^{+0.52} $ for the angular profile of bulk Lorentz factor, and
viewing angle $\protect\theta _{\text{v}}=0.53$ rad. Thus we get $L\sim
10^{47}$ erg s$^{-1}$ and $\Gamma \sim 20$ at the line of sight. For the
model calculation in Section 2.3, we take $L_{0}=10^{50}$ erg s$^{-1},%
\protect\theta _{c,L}=0.1$ rad and $q=3$ for the angular profile of
luminosity, and $\protect\eta _{0}=200$, $\protect\theta _{c,\Gamma }=0.1$
rad and $p=q/4=0.75$ for the angular profile of bulk Lorentz factor. The
viewing angle $\protect\theta _{\text{v}} $ is taken to be $0.8$ rad to get $%
L\sim 10^{47}$ erg s$^{-1}$ and $\Gamma \sim 26$ ($\protect\eta \sim 40$) at
the line of sight. }
\end{figure}

The jet adopted here is a structured jet with an angle-dependent luminosity
(the injected power at the base of the flow) and baryon loading parameter%
\footnote{%
Notice that the baryon loading parameter $\eta $ at the base of the flow is
also the bulk Lorentz factor $\Gamma $ in the saturated acceleration regime.}
outside a uniform core \citep[e.g.,][]{Dai2001,Rossi2002,ZhMe2002,Kumar2003}%
, i.e.,
\begin{eqnarray}
L(\theta ) &=&\frac{L_{0}}{[(\theta /\theta _{c,L})^{2q}+1]^{1/2}},\text{ \
\ \ \ }  \notag \\
\eta (\theta ) &=&\frac{\eta _{0}}{[(\theta /\theta _{c,\Gamma
})^{2p}+1]^{1/2}}+1.2,
\end{eqnarray}%
where $\theta $ is the angle measured from the jet axis, $\theta _{c,L}$ and
$\theta _{c,\Gamma }$ are the half-opening angles for the luminosity core
and the bulk Lorentz factor core ($\theta _{c,L}=\theta _{c,\Gamma }$ is
considered in our calculation), $L_{0}$ and $\eta _{0}$ are corresponding
constant values in the core, respectively, $q$ and $p$ describe how the
luminosity and the bulk Lorentz factor decrease outside the core. Figure 1
presents the shape of the luminosity and Lorentz factor structures and the
best-fit parameters presented in Section \ref{sec:fit}.

\subsection{Photosphere Emission Spectrum}

In the traditional photosphere model, the photospheric radius $R_{\text{ph}}$
is the radius where the scattering optical depth for a photon moving towards
the observer is equal to unity ($\tau =1$). However, one should realize that
wherever there is an electron, a photon has a probability to be scattered
there. For an expanding shell, photons can be last-scattered at any position
in the shell with a probability depending on the position. This changes the
traditional spherical shell photosphere to a probability photosphere
discussed by several authors~\citep{Pe2008,Belo2011,Pe2011,Lund2013,Deng2014}%
. Following the literature, we define a probability function
$P_{1}(r,\Omega )$ as the probability for a photon being last
scattered at the radius $r$ and angular coordinate $\Omega $. This
probability function may be calculated by (see \citealt{Lund2013})
\begin{equation}
P_{1}(r,\Omega )=(1+\beta )D^{2}\times \frac{R_{\text{ph}}}{r^{2}}\exp
\left( -\frac{R_{\text{ph}}}{r}\right) ,
\end{equation}%
where $\beta $ is the jet velocity and $D$ $=[\Gamma (1-\beta \cdot \cos
\theta )]^{-1}$ is the Doppler factor.

In order to obtain the observed spectrum we need to know the probability of
the observer-frame photon energy $E$ when the photon undergoes the last
scattering at $(r,\Omega )$. This photon energy distribution in the observer
frame is determined by that in the co-moving frame and $E=D(\Omega )\cdot
E^{^{\prime }}$, where $E^{^{\prime }}$ is the co-moving frame photon
energy. The photon energy distribution in the local co-moving frame is
assumed to be a Planck function with the same temperature as the electron
due to the coupling of photons and electrons. Then the photon temperature in
the observer frame $T^{ob}$ at $(r,\Omega )$ can be deduced from the plasma
temperature $T^{\prime }(r,\Omega )$ through $T^{ob}=D(\Omega )\cdot $ $%
T^{^{\prime }}(r,\Omega )$. Thus, we can get the distribution function $%
P_{2}(r,\Omega ,E)$ of a photon of energy $E$ and temperature $T^{ob}$ at $%
(r,\Omega )$, which is described as
\begin{equation}
P_{2}(r,\Omega ,E)=\frac{1}{2.40(kT^{ob}(r,\Omega ))^{3}}\frac{E^{2}}{\exp
(E/kT^{ob}(r,\Omega ))-1}.
\end{equation}

\begin{figure*}[th]
\label{Fig_2} \centering\includegraphics[angle=0,height=2.0in]{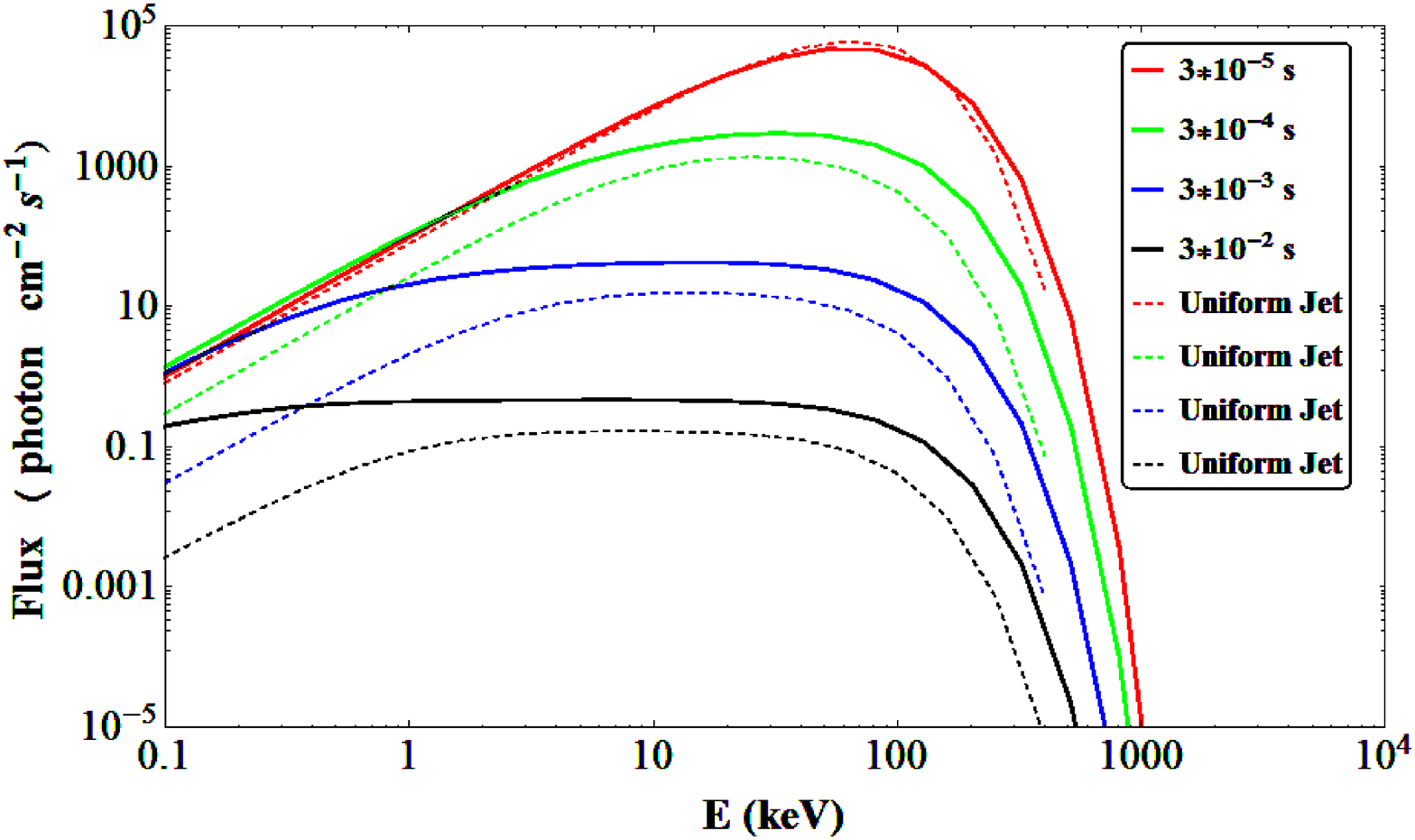} \
\ \centering\includegraphics[angle=0,height=1.95in]{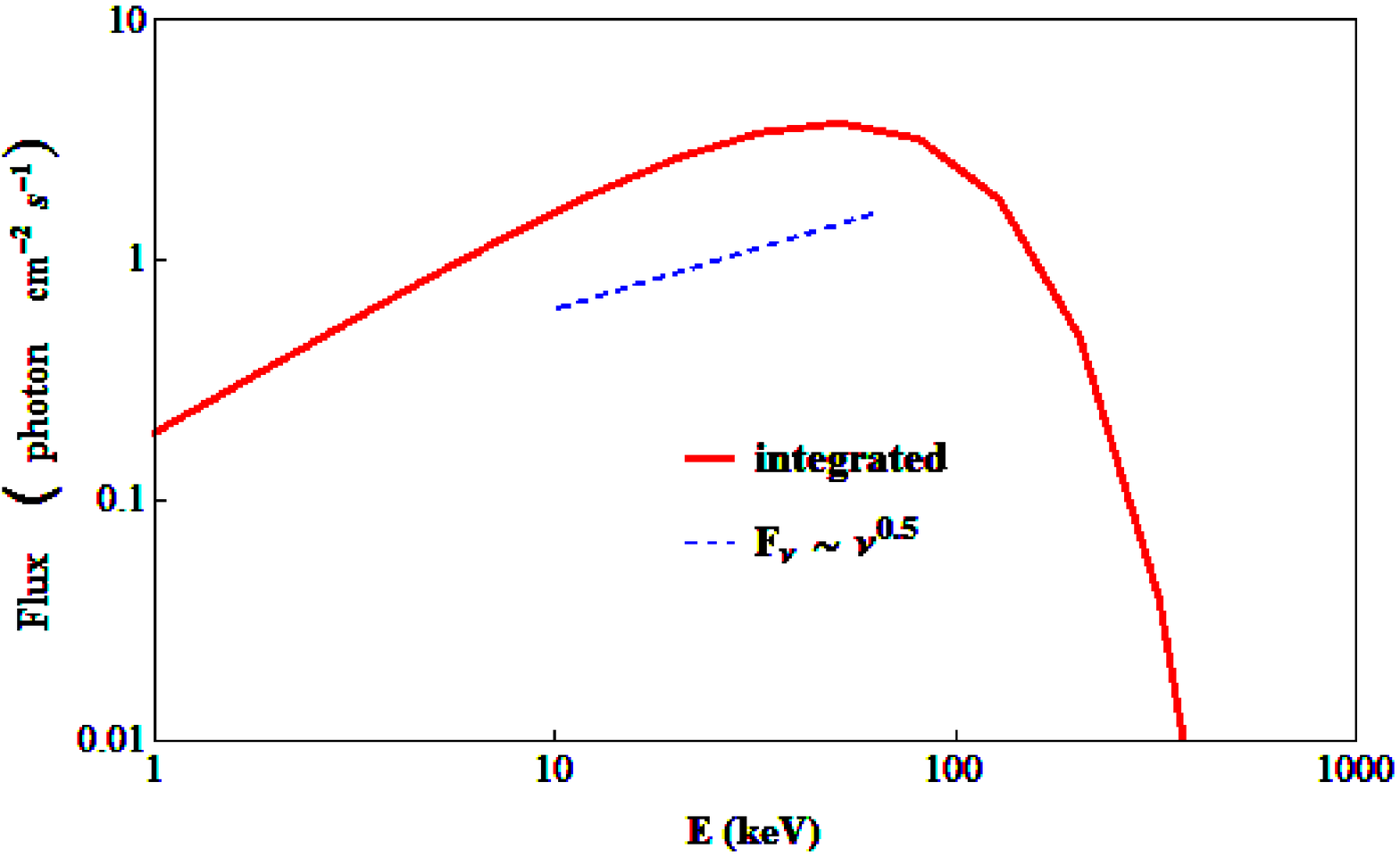} \ \
\caption{The calculated time-resolved spectra and the time-integrated
spectrum. Left panel: \ the solid lines show the time-resolved spectra
calculated with the parameters of the structured jet described in the text.
The dashed lines show the time-resolved spectra calculated in
\citet{Deng2014} for a uniform jet. For the case of a structured jet, the
low-energy flux at later times is greatly boosted. Right panel: the
time-integrated spectrum for the structured jet. The spectrum has a much
softer low-energy photon index $\protect\alpha \sim -0.5$ than blackbody and
an exponential high-energy cutoff, which are close to the empirical fitting
results of the main pulse spectrum of GRB 170817A.}
\end{figure*}

When calculating the observed time-integrated spectrum in the following, we
adopt the spherical coordinates $(r,\Omega (\theta _{_{\text{LOS}}},\phi _{_{%
\text{LOS}}}))$ corresponding to the line of sight (LOS). The observed
time-integrated spectrum is a collection of the photons last scattered at
any position $(r,\theta _{_{\text{LOS}}},\phi _{_{\text{LOS}}})$ and towards
the observer, thus we must know the probability for the last scattering to
occur at \ $(r,\theta _{_{\text{LOS}}},\phi _{_{\text{LOS}}})$ as well as
the temperature at that location. This probability and temperature are
determined by the luminosity and Lorentz factor in the direction $(\theta
_{_{\text{LOS}}},\phi _{_{\text{LOS}}})$, which depend completely on the
angle $\theta $ of this direction to the jet axis. If the angle between the
jet axis and the LOS (i.e. the viewing angle) is $\theta _{v}$, the
corresponding angle $\theta $ follows
\begin{align}
\theta & =\theta (\theta _{_{\mathrm{LOS}}},\phi _{_{\mathrm{LOS}}})  \notag
\\
& =\arccos [\cos (\theta _{_{\mathrm{LOS}}})\cos (\theta _{v})+\sin (\theta
_{_{\mathrm{LOS}}})\sin (\theta _{v})\cos \phi _{_{\mathrm{LOS}}}]\;.
\end{align}%
The time-integrated spectrum can thus be calculated as\footnote{%
Notice that \citet{Deng2014} provided a two-dimensional last scattering
probability function $P(r,\Omega )$. We adopt the separated probability
function $P_{1}$ in this paper, since it is more easily generalized to
structured jets and MCMC fitting.} (see Equation 10 in \citealt{Lund2013})
\begin{equation}
F_{E}^{ob}(\theta _{v})=\frac{1}{4\pi d_{L}^{2}}\iint \frac{d\dot{N}_{\gamma
}}{d\Omega }\times P_{1}(r,\Omega )\times P_{2}(r,\Omega ,E)Ed\Omega dr,
\label{a}
\end{equation}%
where $d\dot{N}_{\gamma }/d\Omega $ is the photon emission rate per unit
solid angle from the base of the outflow ($r=r_{0}$).

In Equation $(\ref{a})$, $d\dot{N}_{\gamma }/d\Omega =(L(\Omega )/4\pi
)/2.7kT_{0}(\Omega )$ , where $L(\Omega )$ is the isotropic luminosity per
unit solid angle $d\Omega $ and $T_{0}(\Omega )=(L(\Omega )/4\pi
r_{0}^{2}ac)^{1/4}$ is the temperature at the base of the outflow per unit
solid angle $d\Omega $. As a result, $d\dot{N}_{\gamma }/d\Omega $ is
angle-dependent.

Since the typical luminosity may be low for a SGRB with rapid decrease of
luminosity in the lateral direction, the photosphere radius $R_{\text{ph}}$
where the photons being last-scattered may be smaller than the saturation
radius for jet acceleration $R_{s}$ =$\eta (\theta )\cdot r_{0}$. We
therefore must judge whether the acceleration is saturated ($R_{\text{ph}%
}>R_{s}$) in each unit solid angle $d\Omega $ by calculating $R_{\text{ph}}$
based on the assumption of saturation, and then deal with them for the
calculations of $P_{1}$ and $P_{2}$ separately. Notice that we have assumed
a pure fireball here for simplicity. In principle, the outflow can be
\textquotedblleft hybrid\textquotedblright\ with an important contribution
from a Poynting flux. The dynamics of such a scenario is more complicated,
but the predicted photosphere spectrum would not be much different from the
pure fireball case, even though the required parameters would be somewhat
different. For a detailed treatment of a hybrid outflow, see \citet{Gao2015}.

For the saturated case, $R_{\text{ph}}$ is given by
\begin{equation}
R_{\text{ph}}=\frac{1}{(1+\beta )\beta \eta ^{2}(\theta )}\frac{\sigma_{T}}{%
m_{p}c}\frac{L(\theta )}{4\pi c^{2}\eta (\theta )},  \label{a1}
\end{equation}%
where $\sigma_{T}$ is the Thompson cross-section, the Doppler factor is $%
D=[\eta (\theta )\cdot (1-\beta (\theta )\cdot \cos \theta _{_{\text{LOS}%
}})]^{-1}$, the observer-frame temperature is $T^{ob}=D(\Omega )\cdot $ $%
T^{^{\prime }}(r,\Omega )$, and the comoving temperature $T^{\prime
}(r,\Omega )$ is
\begin{equation}
T^{\prime }(r,\Omega )=\left\{
\begin{array}{c}
\frac{T_{0}(\Omega )}{2\eta (\Omega )}\text{ \ \ \ \ \ \ \ \ \ \ \ \ \ \ \ \
\ \ \ \ \ },r<R_{s}(\Omega )<R_{\text{ph}}(\Omega )\text{ } \\
\frac{T_{0}(\Omega )\cdot \lbrack r/R_{s}(\Omega )]^{-2/3}}{2\eta (\Omega )}%
\text{ \ \ \ \ \ \ },R_{s}(\Omega )<r<R_{\text{ph}}(\Omega )\text{ } \\
\frac{T_{0}(\Omega )\cdot \lbrack R_{\text{ph}}(\Omega )/R_{s}(\Omega
)]^{-2/3}}{2\eta (\Omega )},R_{s}(\Omega )<R_{\text{ph}}(\Omega )<r\text{ }%
\end{array}%
\right.  \label{b}
\end{equation}

For the unsaturated case, $R_{\text{ph}}$ is calculated by
\begin{equation}
R_{\text{ph}}=\left[ \frac{\sigma _{T}}{6m_{p}c}\frac{L(\theta )}{4\pi
c^{2}\eta (\theta )}r_{0}^{2}\right] ^{1/3}\text{.}
\end{equation}%
In this case, the Lorentz factor at the photosphere and the corresponding
Doppler factor are given by $\Gamma (\theta )=R_{\text{ph}}(\theta )/r_{0}$
and $D=[\Gamma (\theta )\cdot (1-\beta (\theta )\cdot \cos \theta _{_{\text{%
LOS}}})]^{-1}$, respectively, and the comoving temperature is%
\begin{equation}
T^{\prime }(r,\Omega )=T_{0}(\Omega )/[2\Gamma (\Omega )]\text{.}  \label{c}
\end{equation}

To calculate the time-resolved spectra, we add a $\delta $-function $\delta
(t-ru/\beta c)$ to Equation $(\ref{a})$, where $u=(1-\beta (\theta )\cdot
\cos \theta _{_{\mathrm{LOS}}})$. One then has%
\begin{eqnarray}
F_{E}^{ob}(\theta _{v},t) &=&\frac{1}{4\pi d_{L}^{2}}\iint \frac{d\dot{N}%
_{\gamma }}{d\Omega }\times P_{1}(r,\Omega )\times P_{2}(r,\Omega ,E)  \notag
\\
&&\times \delta (t-\frac{ru}{\beta c})Ed\Omega dr.
\end{eqnarray}

With the above analysis, we can derive the time-resolved spectra for
impulsive injection of energy and the time-integrated spectrum for
continuous long-duration energy injection. For a realistic SGRB the duration
for energy injection from the central engine is long ($\sim 1$ s), as
manifested by its observed duration($T_{90}$).

\subsection{Calculated Spectrum}

The parameters of the jet structure and the viewing angle $\theta _{\text{v}%
} $ adopted in our calculation are close to the best-fit values shown in
Figure 1. We set the luminosity at the line of sight to be $\sim $ $10^{47}$
erg s$^{-1}$ to match the observation of GRB 170817A. According to SGRBs
data, typically one has $L_{0}\sim $ $10^{50}$ erg s$^{-1}$ and $\theta
_{c,L}$ $\simeq $ $6^{\circ }-16^{\circ }$ \citep{Fong2015,Ghir2016}. For a
power-law structured jet, the parameter $q$ may be obtained through the
luminosity dependence of local event rate density $\rho _{0}(>L)$ of SGRBs %
\citep[e.g.][]{ZhMe2002}. Since $\rho _{0}(>L)$ $\propto $ $L^{-\lambda }\
(\lambda \sim 0.7,$ \citealt{Sun2015}) and\ $\rho _{0}(>L)$ $\propto $ $%
\Omega (>E)\simeq \pi \theta ^{2}$, the isotropic-equivalent luminosity $L$ $%
\propto $ $\theta ^{-2/\lambda }$ $\propto $ $\theta ^{-q}$, then $q\simeq $
$2.86$. Thus we take $L_{0}=10^{50}$ erg s$^{-1},\theta _{c,L}=0.1$ rad, and
$q=3$ here. Meanwhile we take the viewing angle $\theta _{\text{v}}$ as $0.8$
rad to match the luminosity mentioned above. With this viewing angle and
other parameters we can obtain the approximate model spectrum and thus check
whether we can perform a more detailed MCMC fit for the spectrum of GRB
170817A. Also by comparing with the best-fit parameters (see Sect.3) and the
model spectrum for those best-fit parameters (shown in the bottom left panel
of Fig.3), we can acquire the degree of the change for the parameters
corresponding to different model spectra. As for the bulk Lorentz factor, we
let the value along the line of sight to be in the range of $(20-40)$ in
order to match the peak energy ($\sim $ $100$ keV) of the observed spectrum.
In addition, we take $\eta $ $\propto $ $L^{1/4}$ according to the
statistical results of a large sample of GRBs \citep{Liang2010,Lv2012}.
Finally, we adopt $\eta _{0}=200$, $\theta _{c,\Gamma }=0.1$ rad, and $%
p=q/4=0.75$.

The left panel of Figure 2 shows the calculated time-resolved spectra and
the right panel is the time-integrated spectrum\footnote{%
When calculating results in Fig.2 we do not make use of the best-fit
parameters in Sect.3 but rather use the example parameters, since the
spectrum for the best-fit parameters is presented in the bottom left panel
of Fig.3.}. Comparing the time-resolved spectra of a structured jet (solid
lines in the left panel) and those of a uniform jet (dashed lines in the
left panel), we can see that the low-energy power-law segment below the peak
energy $E_{p}$ is softer than the uniform jet case, and the total fluxes are
also higher. This is because the low-energy emission has a significant
contribution from the high latitudes with respect to the line of sight in
the directions with smaller angles from the jet axis where intrinsic
luminosity is high but Doppler factor is low.

The low-energy photon index is $\alpha \sim -0.5$ for the time-integrated
spectrum in the right panel. This is much softer than the case of the
uniform jet ($\alpha \sim 0.5,$ \citealt{Deng2014}). The origin of such a
difference is again due to the enhanced near-axis high-latitude emission,
likely caused by structures or change in Lorentz factor and luminosity.
There are two effects here. First, the luminosity structure enhances the
near-axis high-latitude emission. Second, the Lorentz factor structure also
allows emission from some directions to become unsaturated, which would also
contribute to the enhancement. The predicted low-energy photon index ($%
\alpha \sim -0.5$) of this model as well as the exponential cutoff on the
high-energy end are consistent with the time-integrated spectrum of GRB
170817A, which can be empirically fitted by a cutoff power-law model with
the low-energy photon index $\alpha \sim -0.6$ \citep{Gold2017,ZhangBB18b}.
This encourages us to perform a more detailed MCMC fit of the data using our
off-axis photospheric emission model from a structured jet.

\section{Spectral Fitting of GRB 170817A with the Off-axis Photosphere Model}

\label{sec:fit}

\begin{figure*}[th]
\label{Fig_3} \centering\includegraphics[angle=0,height=3.0in]{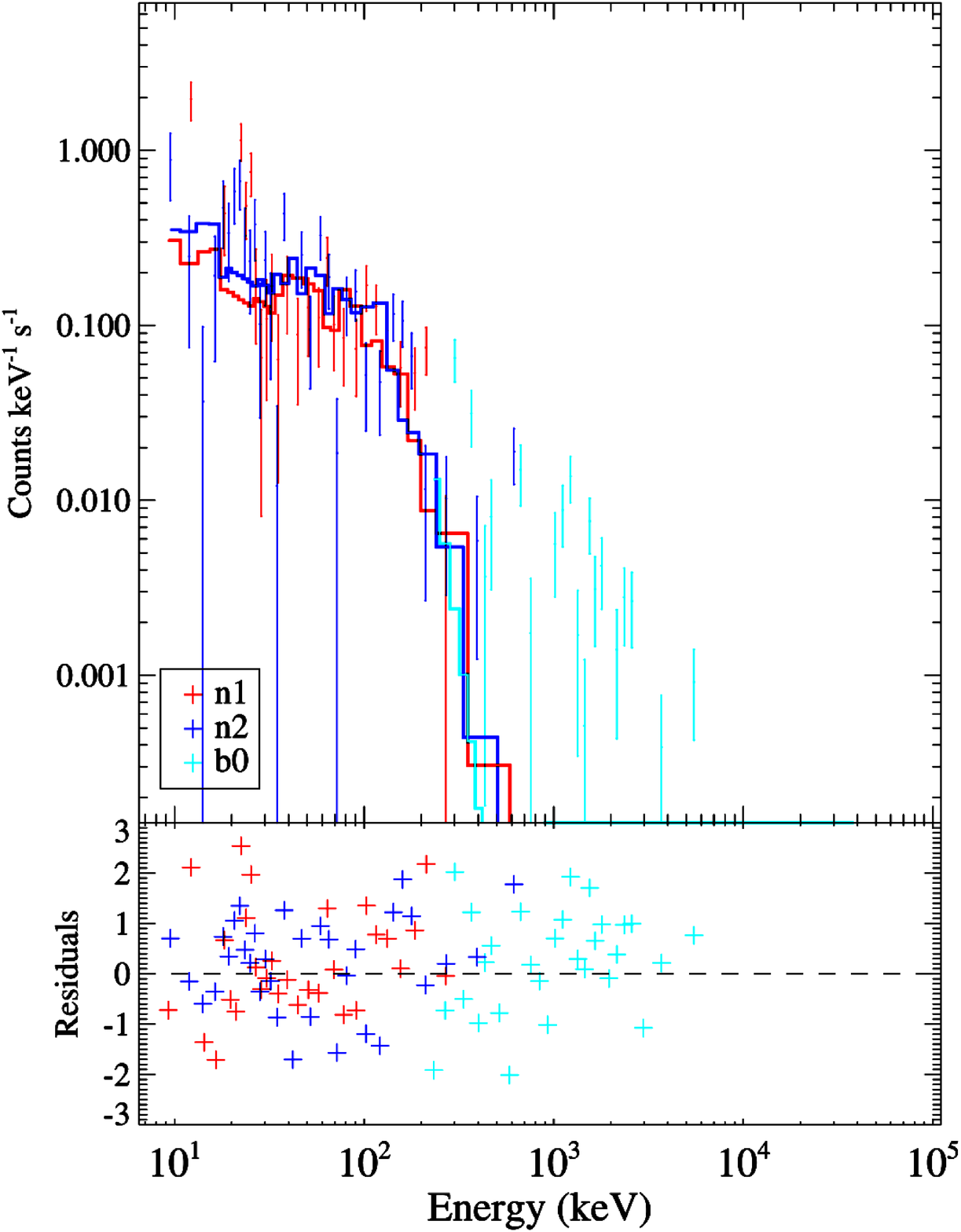}
\ \ \centering\includegraphics[angle=0,height=3.0in]{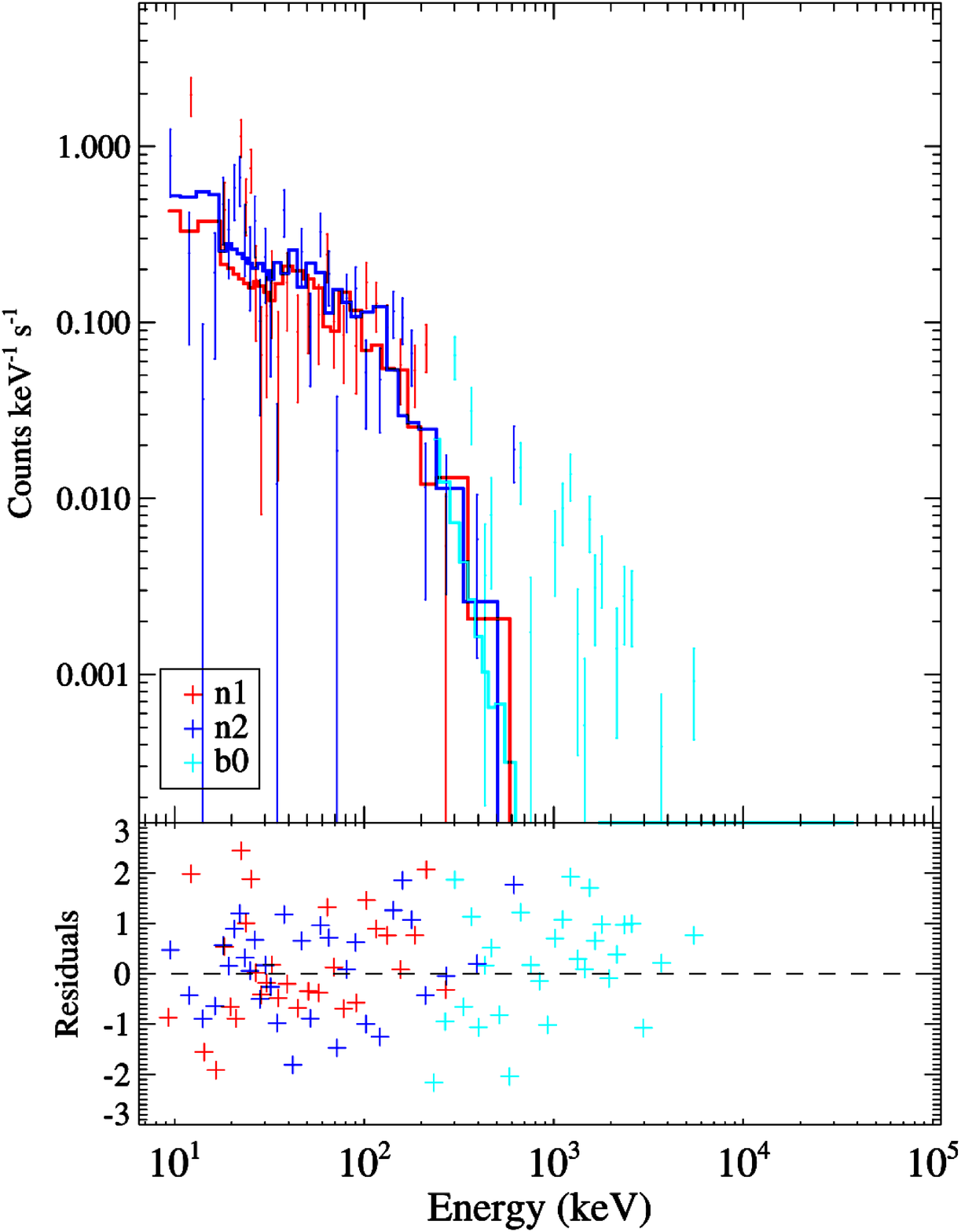} \ \ %
\centering\includegraphics[angle=0,height=3.0in]{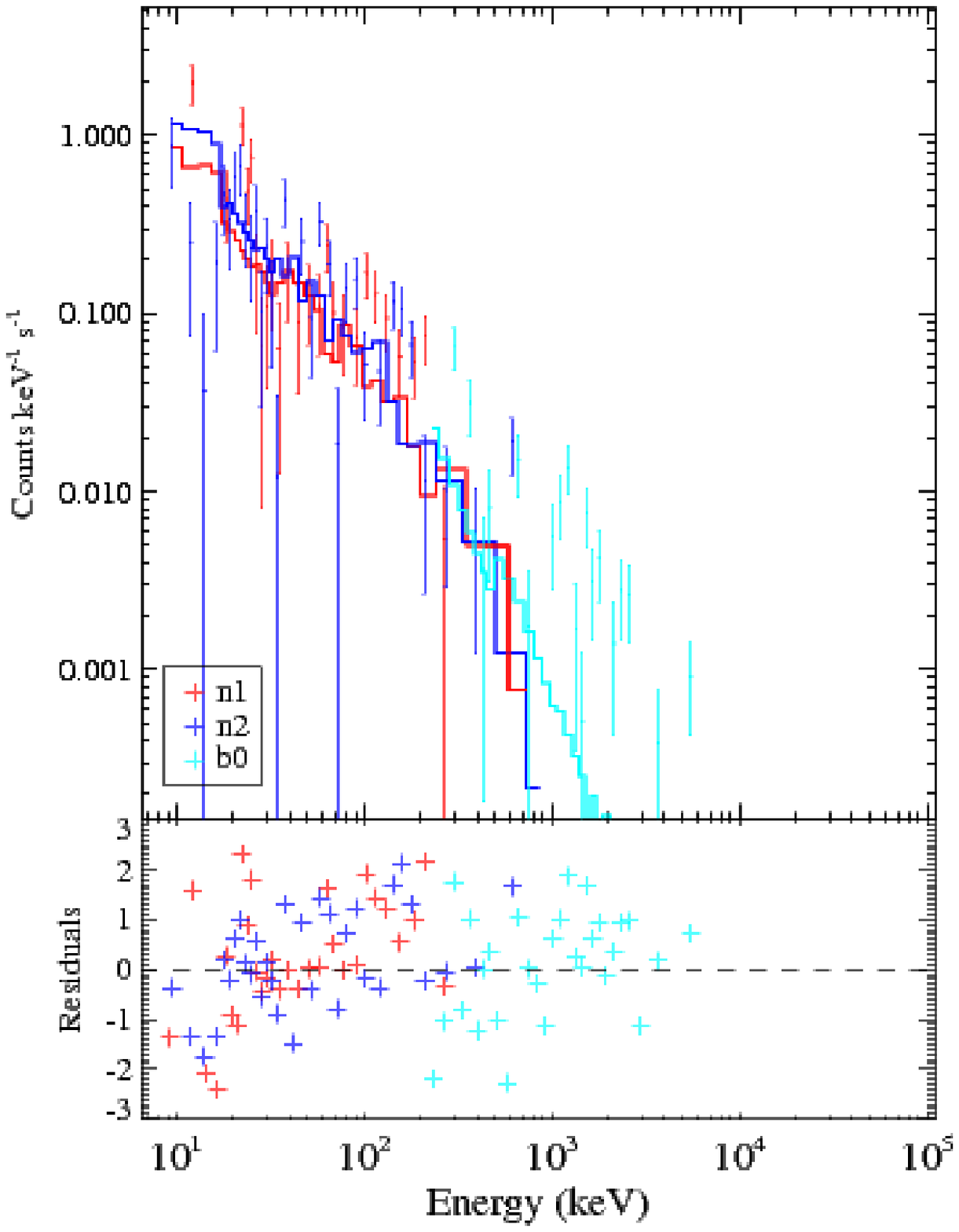} \ \ %
\centering\includegraphics[angle=0,height=2.8in]{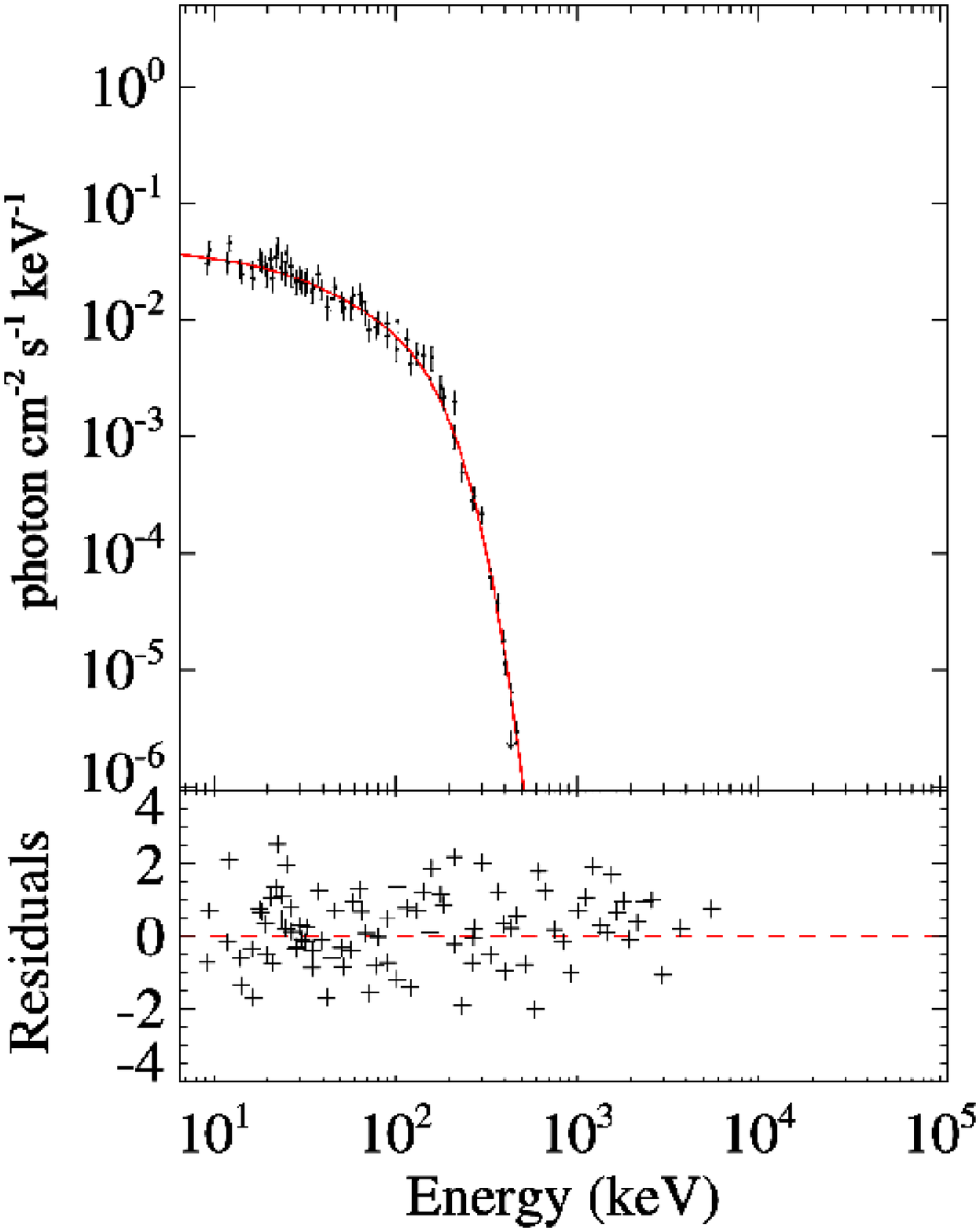} \centering%
\includegraphics[angle=0,height=2.8in]{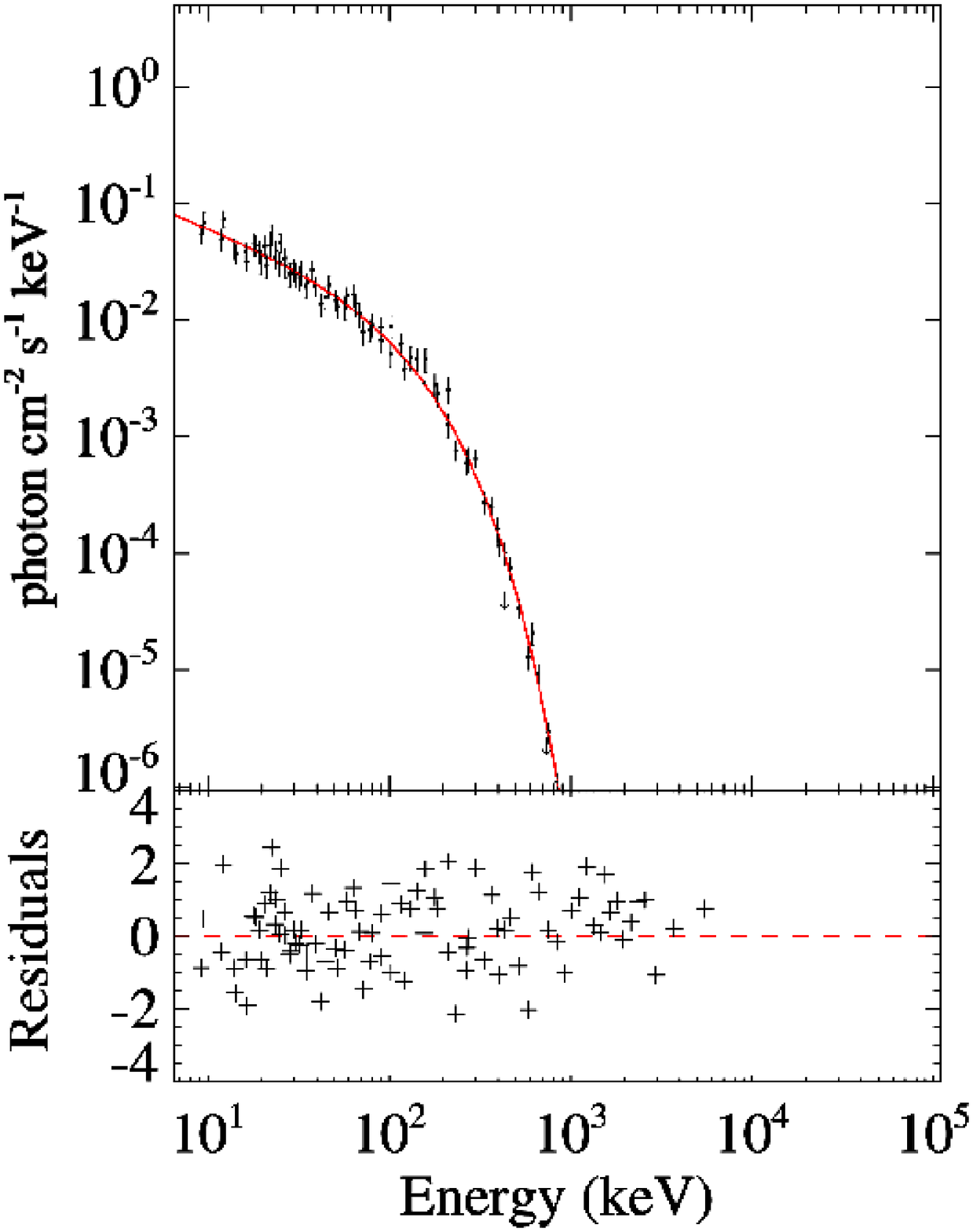} \centering%
\includegraphics[angle=0,height=2.75in]{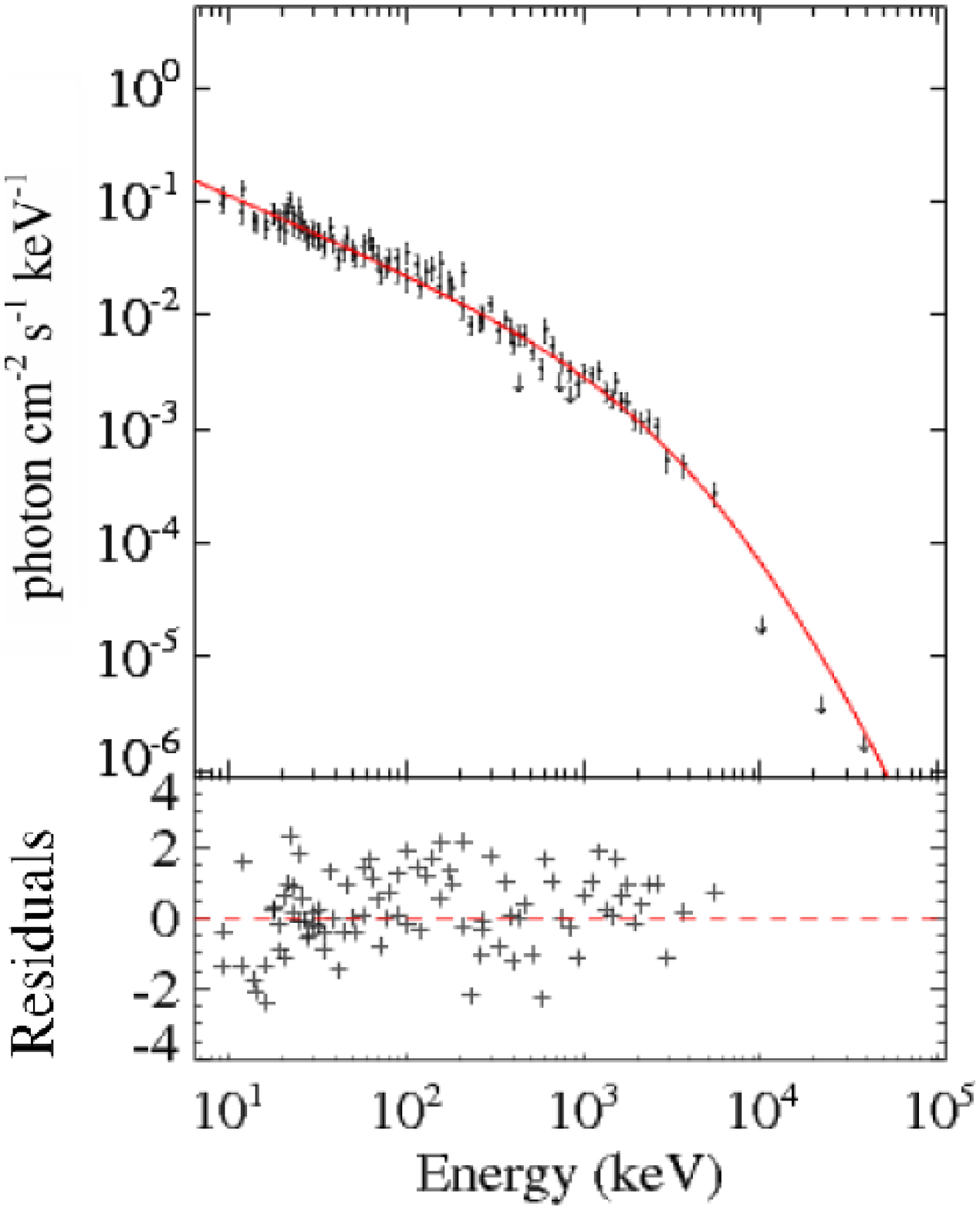} \ \ \ \ \ \ \ \ \ \ \
\ \ \ \ \ \ \ \ \
\caption{Comparisons among our photosphere model fitting, the cutoff
power-law model fitting and the synchrotron model fitting for the
time-integrated spectrum between $-$0.3 s and 0.4 s. Top panels: observed
count spectrum and model count spectrum for our photosphere model fitting
(top left), the cutoff power-law model fitting (top middle) and the
synchrotron model fitting (top right). Bottom panels: theoretical photon
spectrum (red line)\ and\ observed photon flux (data points, which are
obtained by using the instrument responses to de-convolve the observed count
spectrum) for our photosphere model fitting (bottom left), the cutoff
power-law model fitting (bottom middle) and the synchrotron model fitting
(bottom right). The legends of \textquotedblleft n1, n2,
b0\textquotedblright\ in the top panels indicate the two~Thallium activated
Sodium Iodide crystal~detectors, named as NaI n1, NaI n2, and one~Bismuth
Germanate crystal~detector, named as BGO b0.}
\end{figure*}

\begin{figure*}[th]
\label{Fig_4}
\centering\includegraphics[angle=0,height=8.0in]{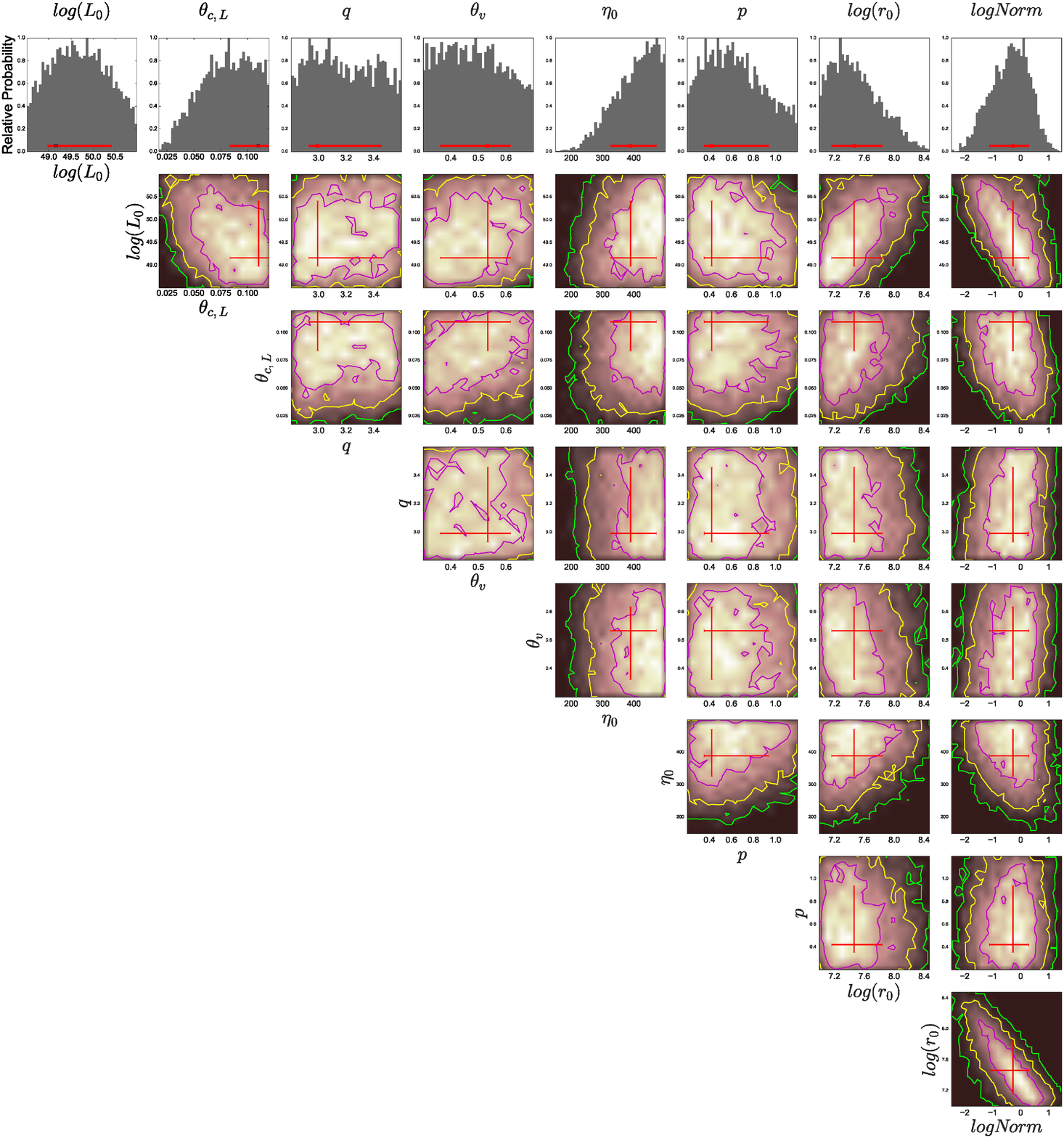} \ \
\caption{Parameter constraints of our photosphere model fitting for
the time-integrated spectrum between $-$0.3 s and 0.4 s. Histograms
and contours illustrate the likelihood map. Red crosses show the
best-fit values and 1-sigma error bars.}
\end{figure*}

GRB 170817A was detected by \textit{Fermi}-GBM and INTEGRAL SPI-ACS, with
the luminosity distance of $\simeq $ 40 Mpc \citep{LIGO2017b}. The analysis
of the \textit{Fermi}-GBM data showed two components: a main pulse from T0$-$%
0.26 s to T0$+$0.57 s and a weak tail extending from T0$+$0.95 s to T0$+$%
1.79 s \citep{Gold2017,ZhangBB18b}. In this work we choose the interval
(i.e., between T0$-$0.3 s to T0$+$0.4 s ) with the most significant emission
to perform the model fitting. We analyze the GBM Time Tagged Event (TTE)
data from detectors NaI 1, NaI 2 and BGO 0. We fit the spectra using our
photosphere model described in Section 2, with McSpecFit package which
accepts flexible user-defined spectral model \citep{ZhangBB16a}. A fit with
the empirical cutoff power-law function was first performed. The spectrum of
this interval is best-fitted by the cutoff power-law model with the
low-energy photon index of $-0.62_{-0.54}^{+0.49}$, peak energy $E_{p}$ = 145%
$_{-26}^{+140}$ keV, and the time-averaged flux of (2.5$_{-1.0}^{+1.8}$) $%
\times $10$^{-7}$ erg cm$^{-2}$ s$^{-1}$. The weak tail between T0$+$0.95 s
and T0$+$1.79 s, with 34\% the fluence of the main pulse, is best fitted by
a blackbody spectrum with $kT=11.3_{-2.4}^{+3.8}$ keV %
\citep{Gold2017,ZhangBB18b}.

A comparison between our photosphere model fitting and the cutoff power-law
model fitting is shown in Figure 3. The best fitting parameters are
presented in Table~\ref{TABLE:MCpho} and also shown in Figure 1. It is
apparent that our photosphere model can fit the data as well as the cutoff
power-law model, with a PGSTAT/dof = 260.9/357 = 0.73 (260.1/363 =0.72 for
the cutoff power-law model). In addition, the residuals do not show any
marked trends.

Parameter constraints of our photosphere model are illustrated in Figure 4.
The best-fit values for the luminosity profile, $L_{0}\sim 10^{49.16}$ erg s$%
^{-1}$, $\theta _{c,L}\sim 0.11$ rad and $q\sim $ $2.99$ are
consistent with the reasonable values of $L_{0}=10^{50}$ erg
s$^{-1},\theta _{c,L}=0.1$ rad and $q=3$
\citep{Fong2015,Sun2015,Ghir2016}. Also, the best-fit values for
the\ bulk Lorentz factor profile, $\eta _{0}\sim
388.82_{-62.9}^{+82.2}$ and $p$ $\sim 0.42_{-0.07}^{+0.52}$ are
close to the reasonable values of $\eta _{0}=200$ and $p=0.75$. The
best-fit viewing angle $\theta _{\text{v}}\sim 0.53_{-0.17}^{+0.08}$
rad falls into the reasonable range ($0.65-0.72$ rad in
\citealt{Gran2017b} and $0.7$ rad in \citealt{Gott2017}). The
observed
luminosity\footnote{%
Since the injected photons are almost emitted at the photosphere, the ratio
of the observed temperature there to the temperature at the base $T_{0}$
represents the efficiency. In the saturated case, the efficiency is ($R_{s}$/%
$R_{\text{ph}}$)$^{2/3}$; while in the unsaturated case the efficiency is $%
\sim $ $1$, which turns out to be the actual case.} at the line of sight is $%
L$ $\simeq 1.3\times $ $10^{47}$ erg s$^{-1}$, which is consistent with the
data \citep{Gold2017,ZhangBB18b}. The best-fit initial radius $r_{0}$ for
acceleration is $\sim 10^{7.46}$ cm. We find that the acceleration is
unsaturated ($R_{\text{ph}}\sim 4.9\times 10^{8}$ cm and $R_{s}\sim $ $%
5\times 10^{9}$ cm) at the line of sight and the actual Lorentz factor%
\footnote{%
Notice that \citet{Zou2018} got a Lorentz factor $\Gamma \sim 13.4$ for the
case of an off beaming relativistic jet.} at the line of sight is $\Gamma
\sim 17$.

The best-fit initial acceleration radius $r_{0}$ is $\sim 10^{7.46}$ cm. %
\citet{Begue2017} gave an estimate of the $r_{0}$ based on the fitted peak
energy and flux of a single blackbody in the observed spectrum (with the
existence of a non-thermal component) using the method of \citet{Pe2007},
and found that $r_{0}$ is too small ($3\times 10^{6}$ cm, close to the
innermost stable circular orbit of a black hole with $3$ $M_{\odot }$) to
justify the photosphere model. This seems to be in contradiction with our
result. We'd like to point out two significant differences between our
photosphere model and theirs. First, the method to estimate the $r_{0}$
given in \citet{Pe2007} is only valid for the case of saturated acceleration
($R_{\text{ph}}>R_{s}$). Thus, the unreasonable low $r_{0}$ only means that
the photosphere model for saturated acceleration is unable to explain the
data well. There is no conflict for our result (large $r_{0}$) since we are
in the unsaturated regime. Second, their method relies on the assumption of
a single blackbody contributed within a small cone along the line of sight,
and an additional non-thermal component is needed to account for the
observed spectrum. Our model, on the other hand, invokes a structured jet so
that emission from high latitudes (relative to the LOS) is included in the
calculation. The resulting spectrum is naturally a multi-color blackbody,
which can account for the observed spectrum well without the need of
introducing a non-thermal component. As a result, our best-fit value $r_{0}$
is justified.

Furthermore, since the acceleration is in the unsaturated regime ($R_{\text{%
ph}}<R_{s}$) along the line of sight, adiabatic cooling is not involved
(unlike the saturated case, see Equation $\ref{b}$ and Equation $\ref{c}$).
As a result, the observed peak energy should be much higher than that in the
saturated case for the same isotropic energy. This seems to be true for this
burst (see Figure 3 in \citealt{ZhangBB18b}).

\begin{table}[tbph]
\caption{Spectral fitting parameters using off-axis photosphere model.}
\label{TABLE:MCpho}\center%
\begin{tabular}{lcl}
\toprule Parameters & GRB 170817A &  \\
\midrule log $L_{0}$ (erg s$^{-1})$ & $49.16_{-0.18}^{+1.25}$ &  \\
$\theta _{c,L}$ (rad) & $0.11_{-0.02}^{+0.01}$ &  \\
$q$ & $2.99_{-0.06}^{+0.46}$ &  \\
$\theta _{\text{v}}$ (rad) & $0.53_{-0.17}^{+0.08}$ &  \\
$\eta _{0}$ & $388.82_{-62.90}^{+82.21}$ &  \\
$p$ & $0.42_{-0.07}^{+0.52}$ &  \\
log $r_{0}$ (cm) & $7.46_{-0.30}^{+0.37}$ &  \\
log Norm & $0.28_{-0.84}^{+0.58}$ &  \\
\bottomrule &  &
\end{tabular}%
\end{table}

\section{Synchrotron Model Fitting}

\label{sec:syn}

\begin{figure*}[h]
\label{Fig_5}
\centering\includegraphics[angle=0,height=8.0in]{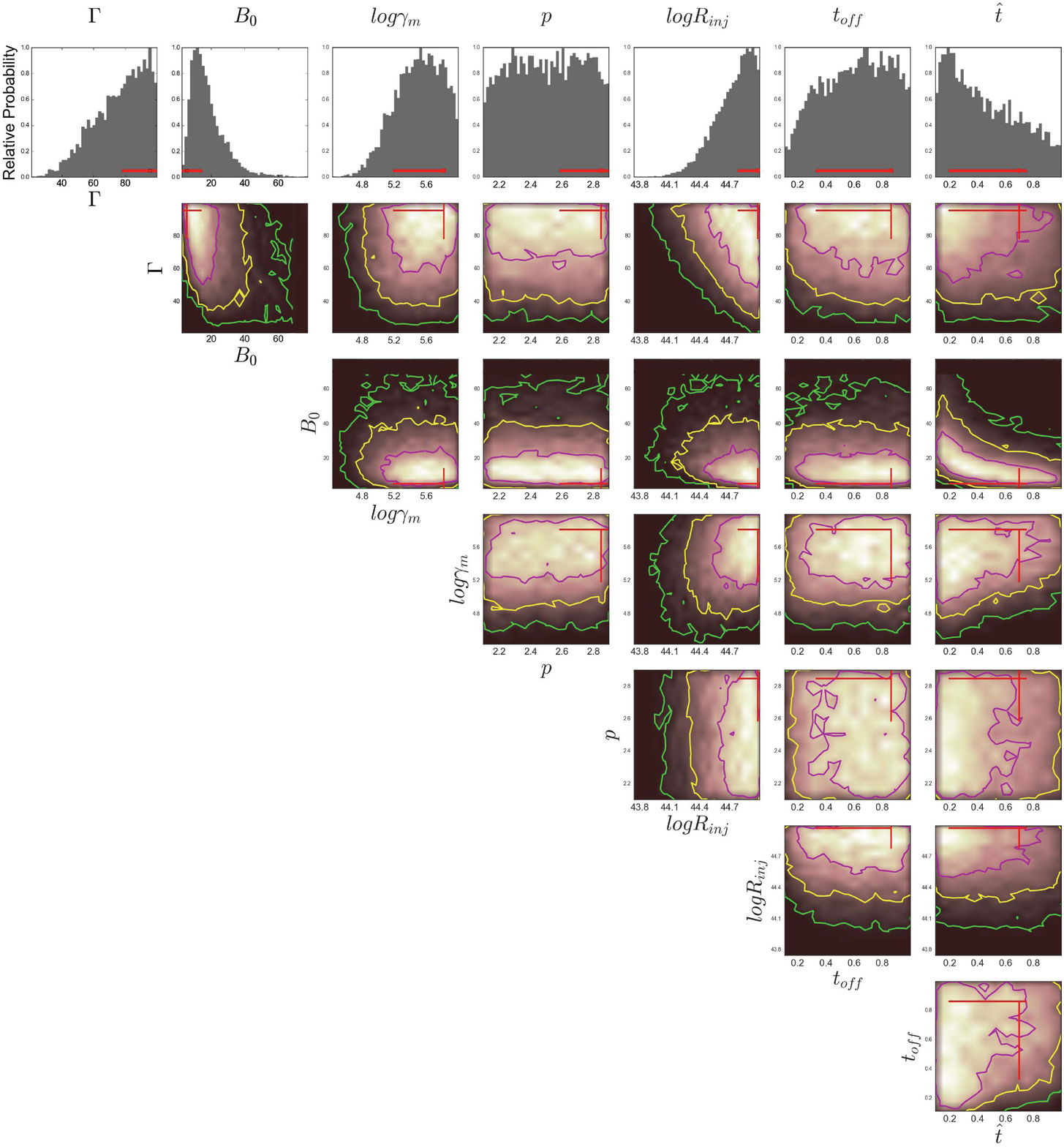} \ \
\caption{Parameter constraints of the synchrotron model fitting for
the time-integrated spectrum between $-$0.3 s and 0.4 s. }
\end{figure*}

Synchrotron radiation from accelerated electrons in an optically thin region
is another promising radiation mechanism for GRB prompt emissions. In this
section, we apply a synchrotron radiation model to fit the spectra of GRB
170817A. To explain the hard low-energy spectrum, \cite{Uhm14} proposed that
fast-cooling electrons in a decaying magnetic field can form a hard electron
distribution, which results in a hard radiation spectrum (also see %
\citealt{Deri2007}). Since the observed spectral index is much harder than
the standard fast-cooling spectrum ($\alpha =-1.5$) \citep{Sari1998}, we
adopt the scenario of synchrotron radiation in a decaying magnetic field %
\citep{Uhm14} in our modeling.

Synchrotron radiation can in principle originate from internal shocks %
\citep{Rees1994} or a magnetic reconnection region (e.g. triggered by
internal-collision-induced magnetic reconnection and turbulence, ICMART) %
\citep{Zhang2011}. The former is relevant for a matter-dominated fireball,
which should be accompanied by a bright photosphere component. If one
interprets the first pulse of GRB 170817A as due to the synchrotron
radiation, the lack of an earlier photosphere component suggests that the
outflow is likely Poynting-flux-dominated, so that the ICMART model may be
more relevant.

Relativistic magnetic reconnection and the shock process are believed to be
able to accelerate non-thermal particles and develop a power-law spectrum of
the particle acceleration (see e.g., \citealt{Guo14,Guo16,Sironi14,Ardaneh15}%
). We assume that a group of electrons, which obey a power-law
distribution, i.e., $Q(\gamma _{\mathrm{e}}^{\prime },t^{\prime
})=Q_{0}(t^{\prime
})(\gamma _{\mathrm{e}}^{\prime }/\gamma _{\mathrm{m}}^{\prime })^{-p}$ for $%
\gamma _{\mathrm{e}}^{\prime }>\gamma _{\mathrm{m}}^{\prime }$, are injected
in the relativistically moving shell of Lorentz factor $\Gamma $. Here, $%
Q_{0}$ is related to the injection rate $N_{\mathrm{inj}}^{\prime }$ by $N_{%
\mathrm{inj}}^{\prime }=\int_{\gamma _{\mathrm{m}}^{\prime }}^{\gamma _{%
\mathrm{max}}^{\prime }}Q(\gamma _{\mathrm{e}}^{\prime },t^{\prime })d\gamma
_{\mathrm{e}}^{\prime }$, where $\gamma _{\mathrm{max}}^{\prime }$ is the
maximum Lorentz factor of electrons. For an electron of $\gamma _{\mathrm{e}%
}^{\prime }$, it would lose energy by synchrotron radiation, of which the
cooling rate is
\begin{equation}
\dot{\gamma}_{\mathrm{e}}^{\prime }=-\frac{\sigma _{T}B^{\prime 2}\gamma _{%
\mathrm{e}}^{\prime 2}}{6\pi m_{\mathrm{e}}c},  \label{eq:syn}
\end{equation}%
where $B^{\prime }$ is the magnetic field in the co-moving frame. Recent
studies reveal that synchrotron self-Compton (SSC) cooling may also play an
important role in shaping the electron energy distribution %
\citep{Bonjak09,Daigne11,Geng18}. However, the effect of SSC cooling on the
resulting spectra is similar to that of decaying magnetic fields. Here we do
not include it for simplicity in our calculations and this would not
markedly impact our main conclusions. Denoting the instantaneous spectrum of
electrons as $\frac{dN_{\mathrm{e}}}{d\gamma _{\mathrm{e}}^{\prime }}$, one
can obtain it by solving the continuity equation in energy space~%
\citep{Longair11}
\begin{equation}
\frac{\partial }{\partial t^{\prime }}\left( \frac{dN_{\mathrm{e}}}{d\gamma
_{\mathrm{e}}^{\prime }}\right) +\frac{\partial }{\partial \gamma _{\mathrm{e%
}}^{\prime }}\left[ \dot{\gamma}_{\mathrm{e}}^{\prime }\left( \frac{dN_{%
\mathrm{e}}}{d\gamma _{\mathrm{e}}^{\prime }}\right) \right] =Q(\gamma _{%
\mathrm{e}}^{\prime },t^{\prime }).  \label{eq:continuity}
\end{equation}

Considering a conical jet, the co-moving magnetic field in the jet would
decay with radius as
\begin{equation}
B^{\prime }=B_{0}^{\prime }\left( \frac{R}{R_{0}}\right) ^{-1},
\end{equation}%
where $B_{0}^{\prime }$ is the magnetic strength at $R_{0}$, and $R_{0}$ is
the radius where the jet begins to emit the first photon observed by us. In
our modeling, we take $R_{0}=2\Gamma ^{2}c\times 1~\mathrm{s}$, and denote
observer-frame time since the first electron injection as $\hat{t}$ (in
units of s) for an emission episode. We further introduce a parameter $t_{%
\mathrm{off}}$ to describe when the injection of electrons is turned off in
the observer frame. Therefore, seven parameters in total are left free,
i.e., $\Gamma $, $\gamma _{\mathrm{m}}^{\prime }$, $B_{0}^{\prime }$, $p$, $%
N_{\mathrm{inj}}^{\prime }$, $t_{\mathrm{off}}$ and $\hat{t}$. Unlike the
calculation method for spectra adopted in Section 2, we only consider the
emission from the region just near the LOS and treat this small region as a
uniform jet. So relevant parameters in our calculation describe properties
of the region near the LOS, rather than those of the jet axis. This
treatment enables us to simplify the calculation and focus on properties of
the region near the LOS. Unlike photosphere emission for which one has
considered the shape of the last-scattering surface which could be
noticeably different for a structured jet, the synchrotron model is not
affected by the jet structure if the Lorentz factor along the LOS is large
enough \citep[e.g.][]{ZhMe2002}. This is valid for our case (our best-fit $%
\Gamma \sim 96$ along the LOS, so our simplification does not impact final
results significantly).

We fit the spectra by interpolating our synchrotron model into the McSpecFit
package (also see \citealt{ZhangBB18a,ZhangBB18b} for details), and the
fitting results are shown in Table~\ref{TABLE:MCsyn} and Figure~5, with a
PGSTAT/dof = 269.4/359. Compared with the PGSTAT/dof = 260.9/357 for the
photosphere model, the PGSTAT/dof for the synchrotron model is slightly
larger. However, this small difference could not help to prefer one model
over the other.

One can perform a self-consistency check of the synchrotron model
parameters. The GRB emission is delayed by $\Delta t \sim 1.7$ s with
respect to the gravitational wave merger time \citep{LIGO2017b,ZhangBB18b}.
If one assumes that the jet is launched right after the merger, the distance
the jet traveled at the time of magnetic dissipation is $R_{\mathrm{GRB}}
\sim \Gamma^2 c \Delta t \sim 4.7\times 10^{14}$ cm. Given the observed
luminosity $L \sim 10^{47} \ \mathrm{erg \ s^{-1}}$, the co-moving magnetic
field in the emission region may be estimated as \citep[e.g.][]{ZhMe2002b} $%
B^{\prime }\leq (2L/c R_{\mathrm{GRB}}^2)^{1/2}/\Gamma \sim 58$ G. The
best-fit parameter falls within this range, suggesting the consistency of
the model.

\begin{table}[tbph]
\caption{Spectral fitting parameters using synchrotron model.}
\label{TABLE:MCsyn}\center%
\begin{tabular}{lcl}
\toprule Parameters & GRB 170817A &  \\
\midrule$\Gamma $ & $95.57_{-17.51}^{+4.43}$ &  \\
$B_{0}^{\prime }$ (G) & $5.45_{-2.76}^{+8.96}$ &  \\
$\mathrm{log}$~$\gamma _{m}^{\prime }$ & $5.82_{-0.63}^{+0.001}$ &  \\
$p$ & $2.85_{-0.26}^{+0.05}$ &  \\
$\mathrm{log}$~$R_{\mathrm{inj}}$ (s$^{-1}$) & $44.98_{-0.20}^{+0.02}$ &  \\
$t_{\mathrm{off}}$ (s) & $0.86_{-0.54}^{+0.01}$ &  \\
$\hat{t}$ (s) & $0.70_{-0.51}^{+0.05}$ &  \\
\bottomrule &  &
\end{tabular}%
\end{table}

Our results suggest that the synchrotron model can also give a reasonable
interpretation to the first pulse of the prompt emission of GRB 170817A.
More complicated effects such as SSC \citep{Geng18} and slow
heating/acceleration for electrons \citep{Xu17,Xu18} have not been
considered in our calculation. However, since these effects also tend to
harden the spectrum, including them would also give a reasonable
interpretation to the data, even though the best-fit parameters may be
somewhat changed.

\section{DISCUSSION}

\label{sec:dis}

\subsection{The Blackbody in the Weak Tail}

The spectrum of the weak tail emission of GRB 170817A is consistent with
being a blackbody. Within our structured jet photosphere model, this may be
interpreted as the transition from a structured jet to a roughly uniform jet
at late times or the change of Lorentz factor and luminosity such that the
contributions to observed flux from high latitudes are weakened. The softer
peak energy is a natural result from the decrease of the luminosity and the
Lorentz factor at late times. According to the best-fit results for the main
pulse above, we have $L$ $\sim 10^{47}$ erg s$^{-1}$, $\eta $ $\sim $ $%
50-150 $ at the line of sight. Thus for the weak tail with $L\sim 0.3\times
10^{47}$ erg s$^{-1}$, if the bulk Lorentz factor $\eta \sim 20$ (saturated
acceleration with $R_{\text{ph}}\sim 3.3\times 10^{9}$ cm and $R_{s}\sim
5.8\times 10^{8}$ cm), we may get a blackbody spectrum with kT=$%
11.3_{-2.4}^{+3.8}$ keV. One should note that these are the average values
within the entire duration of the weak tail.

Within the synchrotron model, the blackbody tail emission should be
attributed to a different mechanism. One may suppose that a successful
structured jet breaks out to make the first pulse via synchrotron, and the
more isotropic component breaks out the cocoon later to make the second
thermal tail. Therefore, it is unable to rule out the synchrotron model
based on the existence of the thermal tail.

\subsection{The Time Delay between the GW Signal and the SGRB}

The $\gamma $-ray emission onset of GRB 170817A has a delay of
$\Delta t$ = 1.74 $\pm $0.05 s relative to the GW chirp signal
\citep{LIGO2017b}. Under the framework of photosphere model, some
additional mechanism is required to account for such a delay. For
instance, this delay may be attributed to the existence of a
short-lived (t$_{\text{HMNS}}$ $\lesssim $ 1 s) hypermassive NS
(HMNS) after the NS$-$NS merger, and the jet is launched only after
the hypermassive NS collapses into a black hole
\citep[e.g.][]{Gran2017}. Such type of the NS$-$NS merger remnant is
supported by previous numerical studies
\citep[e.g.,][]{Ross2002,Ross2003}. The delay onset of a
relativistic jet relative to the merger is also required by the cocoon model %
\citep[e.g.][]{Gott2017}. After launching, the relativistic jet needs to
break through the dynamical ejecta \citep[e.g.,][]{Hoto2013,Ross2013} and/or
neutrino driven wind, causing another time delay that could be a large
fraction of a second \citep[e.g.,][]{Moha2017,Nakar2017}.

Within the photosphere model, if one assumes $\Gamma \approx 2-3$ along the
line of sight for the structured jet, the observed delay can be well
explained without introducing an extra delay for the onset of the jet. In
this case, however, the photosphere temperature is too low to explain the
observed $E_{p}$. One needs to introduce some sub-photospheric dissipative
processes to boost up $E_{p}$ through Comptonization %
\citep{Ree2005,Gian2006,Begue2015,Vur2016}.

Within the synchrotron model, one does not need to invoke such a delayed
launch of jet with respect to the merger time. The delay can be accounted
for by the time scale when the relativistic jet reaches the dissipation
radius. It is intriguing that both the duration of the burst and the delay
time are of the same order. Within the synchrotron model, both time scales
are related to $R_{\mathrm{GRB}}/c\Gamma ^{2}$, and therefore are comparable %
\citep{ZhangBB18b}.

\subsection{Comparison with the Cocoon Emission Model}

Using the cocoon shock breakout to explain the $\gamma $-ray
emission of GRB 170817A has been proposed lately
\citep[e.g.,][]{Gott2017,Kas2017,Brom2018}. A delayed launch of the
jet after the merger is needed to explain the data. In order to
explain the soft low-energy photon index of the main pulse spectrum,
both the cocoon shock breakout and our scenario attribute the soft
emission below $E_{p}$ to the superposition of a series of blackbody
with different temperatures. The significant difference between
their model and ours is the origin of low luminosity. In our model,
the low luminosity is caused by the low luminosity of the structured
jet along the line of sight, since we think that the jet may have a
decreasing luminosity with angle and the viewing angle is large. The
low luminosity of the cocoon shock breakout
model arises from the low mass (thus low internal energy, $m_{\text{tail}%
}\sim $ $4\times 10^{-7}$ $M_{\odot }$) of the fast ejecta tail which emits $%
\gamma $-ray photons with a small Lorentz factor $\Gamma _{s}\approx 2-3$.

It is worth emphasizing that GRB 170817A appears a natural extension of
short GRBs to the low-luminosity regime. The duration ($T_{90}$) and the
peak energy of GRB 170817A are similar to a group of short GRBs %
\citep{Lu2017,ZhangBB18b}. The average low-energy photon index ($\alpha $ $%
\sim -$0.69, \citealt{Bur2017,Lu2017}) for the complete short GRBs sample of
\textit{Fermi} GBM is close to the low-energy photon index ($\alpha $ $\sim
- $0.62) of this burst. The SGRB event rate density above a much lower
luminosity threshold ($\sim 10^{47}$ erg s$^{-1}$), obtained by including
GRB 170817A, is found to be consistent with the extension of the PL
distribution for the normal SGRBs with higher luminosities \citep{ZhangBB18b}%
. All these suggest that GRB 170817A may not have a very different origin
from other short GRBs. The radiation mechanism for GRB 170817A is likely to
be the same as that of other short GRBs with high luminosity. We believe
that photosphere emission or synchrotron radiation from a structured jet
with a large viewing angle is a natural explanation to the prompt emission
data of GRB 170817A, and the cocoon model may not be needed to account for
the data\footnote{%
We stress that the cocoon may still exist in our models. But for our
scenarios the outflow from the central engine can break out the
cocoon quickly and naturally develop a structured jet, which is
ahead of the slowly expanding cocoon. Further studies and detailed
numerical simulations are needed to test this possibility.}. It has
been suggested that the recent brightening of radio and X-ray fluxes
is consistent with the prediction of the cocoon model
\citep{Kas2017}. On the other hand, the structured jet model can
also explain the same data available so far \citep{Lazza2017b} as
well as the late-time optical afterglow \citep{Lyman2018}.

\section{CONCLUSIONS}

\label{sec:con}

As the first short GRB detected to be associated with a NS$-$NS merger
event, GRB 170817A carries important clues to unveil the underlying physics
of SGRBs, including jet launching, interaction with the dynamical ejecta,
energy dissipation mechanism, and radiation mechanism. The prompt emission
data can be used to constrain these mechanisms.

In this paper, we focus on the spectral data of the first emission episode
of GRB 170817A, and explore two models to account for the observed data. We
find that both models can give reasonable fit to the data. In the first
model, we developed a photosphere model in a structured jet. We found that
the emission from the part closer to the jet axis can enhance the low-energy
component of the spectrum, resulting in a softer low-energy photon index ($%
\alpha \sim -0.5$) which is consistent with the observation ($\alpha \sim
-0.6$). We performed a MCMC fit of the spectrum from T0 $-$ 0.3 s to T0$+$%
0.4 s using our model, and found that our model can give a comparable fit to
the best-fit empirical model (the cutoff power-law model). The best-fit
parameters are consistent with the results from some statistic works for
SGRBs. In the second model, we consider synchrotron radiation in an
optically thin region with the jet expanding with a decaying magnetic field
strength. This model also gives a reasonable fit to the data, even though a
higher Lorentz factor along the line of sight is needed.

GRB 170817A is observed to be delayed from GW170817 by $\sim 1.7$ s. Within
the photosphere model, one needs to introduce a delay of the launch of the
jet after the merger. Such a requirement is also needed by the cocoon shock
break model. The synchrotron model does not demand such a delay time.

\cite{Begue2017} discussed whether the typical emission models of
synchrotron radiation and photospheric emission for structured and top-hat
jets can explain the prompt emission of GRB 170817A, and found that these
models are particularly challenging. They then proposed that the standard
models for SGRBs need to be modified. We reached an opposite conclusion by
introducing a structured jet so that the observed spectrum is intrinsically
multi-color blackbody. Another difference is that jet acceleration is in the
unsaturated regime. As we have shown, the photosphere model can give a very
good fit to the data. For synchrotron radiation, we reached a set of
best-fit parameters which are not unreasonable, in contrast to the
conclusion of \cite{Begue2017}. We therefore conclude that both mechanisms
are not ruled out by the data, and that the standard GRB mechanism (with a
large viewing angle to a structured jet) can account for the prompt emission
data of GRB 170817A without the need to invoke a different mechanism, e.g.
cocoon shock breakout.

\acknowledgments

We thank the referee for helpful suggestions. We acknowledge the use of the
public data from the \textit{Fermi} data archives. This work is supported by
the National Basic Research Program (\textquotedblleft
973\textquotedblright\ Program) of China (Grant No. 2014CB845800), the
National Natural Science Foundation of China (Grant Nos. 11725314, 11673068,
11433009, 11543005, 11603076, 11573014, 11533003, 11473012, 11722324,
11603003, 11633001 and 11690024), the Youth Innovation Promotion Association
(2011231 and 2017366), the Key Research Program of Frontier Sciences
(QYZDB-SSW-SYS005), the Strategic Priority Research Program
\textquotedblleft Multi-waveband gravitational wave
Universe\textquotedblright\ (Grant No. XDB23000000) of the Chinese Academy
of Sciences, and the Natural Science Foundation of Jiangsu Province (Grant
No. BK20161096). JJG is supported by the National Postdoctoral Program for
Innovative Talents (Grant No. BX201700115) and China Postdoctoral Science
Foundation funded project (Grant No. 2017M620199). BBZ acknowledge the
support from the National Thousand Young Talents program of China.

\end{document}